\documentclass[10pt,aps,twocolumn,nobibnotes,pra,showpacs,posprint]{revtex4-1}
  \usepackage{color}
  \usepackage{newlfont}
  \usepackage{graphicx}
  \usepackage{amssymb}
  \usepackage{amsmath}
  \usepackage{bm}
  \usepackage{verbatim}
  \usepackage[latin1]{inputenc}
  \usepackage{bbm}
 \usepackage[varg]{txfonts}

\newcommand{\ket}[1]{\mbox{$ | #1 \rangle $}}
\newcommand{\bra}[1]{\mbox{$ \langle #1 | $}}
\newcommand{\be}{\begin{equation}}
\newcommand{\ee}{\end{equation}}
\newcommand{\beq}{\begin{eqnarray}}
\newcommand{\eeq}{\end{eqnarray}}
\begin{document}

\title{Classical-hidden-variable description for entanglement dynamics of two-qubit pure states}
\author{L. S. Silveira}
\author{R. M. Angelo}
\affiliation{Department of Physics, Federal University of Paran\'a, P.O.Box 19044, 81531-980, Curitiba, PR, Brazil}

\begin{abstract}
A hidden-variable model is explicitly constructed by use of a Liouvillian description for the dynamics of two coupled spin-1/2 particles. In this model, the underlying Hamiltonian trajectories play the role of deterministic hidden variables, whereas the shape of the initial probability distribution figures as a hidden variable that regulates the capacity of the model in producing correlations. We show that even though the model can very well describe the short-time entanglement dynamics of initially separated pure states, it is incapable of violating the Clauser-Horne-Shimony-Holt inequality. Our work suggests that, if one takes the reluctance of a given quantum resource to be emulated by a local-hidden-variable model as a signature of its nonclassicality degree, then one can conclude that entanglement and nonlocality are nonequivalent even in the context of two-qubit pure states.
\end{abstract}


\maketitle

\section{Introduction}

In response to Einstein, Podolsky, and Rosen's 1935 paper~\cite{EPR35} claiming the incompleteness of quantum mechanics, Bell showed that no eventual theory compatible with the notion of local causality can fully reproduce the results of quantum theory~\cite{bell64}. Recently, loophole-free Bell's tests~\cite{hensen15,giustina15,shalm15,hensen16} have shown that quantum mechanics is in perfect consistency with nature, which is then believed to admit some sort of nonlocal aspects while forbidding signaling.

Entanglement plays a prominent role in this scenario~\cite{horodecki09}. Interpreted as a class of quantum correlation that cannot be created via local operations and classical communication, entanglement is, in general, necessary but not sufficient for the manifestation of Bell nonlocality~\cite{wiseman07,brunner14,quintino15,costa16}. (Recently, however, it has been shown that some nonlocal aspects can be found even in the absence of entanglement~\cite{bilobran15}.) As far as pure states are concerned, Gisin's theorem~\cite{gisin91} and its most general form~\cite{yu12} establish that all pure states violate a single Bell inequality, thus proving equivalence between entanglement and Bell nonlocality (hereafter referred to simply as nonlocality).

Motivated by the need of a better understanding of the relation between entanglement and nonlocality, researchers have investigated scenarios involving local entanglement. It has been shown that, in the context of nonsequential local measurements, there exist genuinely multipartite entangled states that admit a local-hidden-variable model~\cite{bowles16}. Also, methods have been proposed to accomplish the rather challenging task of constructing local-hidden-variable models for general quantum states~\cite{cavalcanti16,hirsch16}.

This work aims at contributing to this context, but within a dynamical perspective. We explicitly construct a hidden-variable model and demonstrate its adequacy in describing pure-state entanglement in the short-time dynamics of two coupled spin-1/2 particles despite the proven local character of the involved hidden variables. This result turns out to be somewhat puzzling in light of Gisin's theorem: since all two-qubit entangled pure states violate the Clauser-Horne-Shimony-Holt (CHSH) inequality, which implies the aforementioned entanglement-nonlocality equivalence, then we have at hand a scenario in which the dynamics of nonlocality can be tracked with a local model. A fundamental result of this paper is to show that, in spite of this peculiar ability of our model, it does not stand against Bell's theorem.

To devise our model, we get some inspiration from previous works on the dynamics of two coupled oscillators initially prepared in a product of coherent states~\cite{angelo04,angelo05}. These works have shown that (i) the generation of entanglement at short times can be faithfully reproduced with the Liouville theory and (ii) pure-state entanglement, as quantified by the linear entropy of a partition, persists (actually, increases) as we move towards the classical limit $(\hbar\to 0)$. These results were posteriorly corroborated by similar studies on different systems~\cite{matzkin11,casati12}. In the context of the present work, instead of figuring as a semiclassical approximation to quantum mechanics, the Liouville theory is to be interpreted as an epistemic statistical theory provided with deterministic hidden variables (trajectories), which is intended to be an alternative to quantum mechanics in describing the world. In this sense, its role is closer to that of Bohmian's interpretation~\cite{bohm52}.

This paper is structured as follows. In Sec.~\ref{CHSH}, we review the CHSH inequality in some detail, emphasizing that it applies to any bounded observables. The equivalence between entanglement and nonlocality is quantitatively illustrated in Sec.~\ref{QM} for the scenario of interest, namely, the dynamics of two-qubit systems prepared in a product of spin coherent states and subjected to bilinear couplings. Section~\ref{CHVM} is devoted to the presentation of our classical-hidden-variable model (CHVM). Besides explicitly constructing deterministic hidden variables and implementing statistical aspects in the model, we introduce a convenient measure of inseparability in clear analogy with entanglement. We then submit our model to a Bell test and, in Sec.~\ref{conclusion}, close the paper with a brief summary and discussions.

\section{Revisiting the CHSH inequality}
\label{CHSH}

Consider the usual Bell scenario involving two spacelike separated observers, Alice and Bob, who can measure generic observables $\mathrm{A}$ and $\mathrm{B}$ and get respective outcomes $a$ and $b$. In general, experimental results are shown not to obey {\em factorizability} (statistical independence); that is, the joint probability is such that $p(a,b|\mathrm{A,B})\neq p(a|\mathrm{A})p(b|\mathrm{B})$. We then look for a model that, supplemented with a hidden variable $\lambda$, can restore factorizability for space-like separated measurements. In this model, the measured joint probability is to be confronted with $p(a,b|\mathrm{A,B})=\int\mathrm{d}\lambda\,p_{\lambda}\,p(a,b|\mathrm{A,B},\lambda)$, with $p_{\lambda}$ being a probability distribution for $\lambda$ satisfying $p_{\lambda}\geqslant 0$ and $\int\mathrm{d}\lambda p_{\lambda}=1$. From the definition of conditional probability, it follows that
\be\label{conditional}
p(a,b|\mathrm{A,B},\lambda)=p(a|b,\mathrm{A,B},\lambda)p(b|\mathrm{A,B},\lambda).
\ee 
Following Ref.~\cite{cavalcanti12}, we now make two crucial assumptions about the model. The first one ({\em determinism}) requires that 
\be\label{determinism}
p(a,b|\mathrm{A,B},\lambda) \in \{0,1\},
\ee
which implies that $a=a(\mathrm{A,B},\lambda)$ and $b=b(\mathrm{A,B},\lambda)$, that is, the outcomes $a$ and $b$ are fully determined by the parameters that define the physical scenario in question. By this assumption, one can replace $p(a|b,\mathrm{A,B},\lambda)$ in Eq.~\eqref{conditional} with $p(a|\mathrm{A,B},\lambda)$ since this probability distribution must not change by conditioning on $b$ ($a$ is already determined by $\{\mathrm{A,B},\lambda \}$). The second crucial assumption ({\em locality}) is expressed as
\be\label{locality}
\begin{array}{l}
p(a|\mathrm{A,B},\lambda)=p(a|\mathrm{A},\lambda),\\   
p(b|\mathrm{A,B},\lambda)=p(b|\mathrm{B},\lambda).
\end{array}
\ee 
With assumptions \eqref{determinism} and \eqref{locality}, Eq.~\eqref{conditional} reduces to
\be 
p(a,b|\mathrm{A,B},\lambda)=p(a|\mathrm{A},\lambda)p(b|\mathrm{B},\lambda),
\ee 
which clearly respects factorizability. The measured probability distribution is then described by this model as
\be\label{Blocality}
p(a,b|\mathrm{A,B})=\int\mathrm{d}\lambda\,p_{\lambda}p(a|\mathrm{A},\lambda)p(b|\mathrm{B},\lambda).
\ee
This expression is usually referred to in the literature as the hypothesis of {\em local causality}, or simply {\em locality}~\cite{brunner14}.

To derive the CHSH inequality, we first note that a generic expectation value is computed in this local-hidden-variable model as
\be\label{<AB>}
\langle \mathrm{A\otimes B}\rangle \equiv\int\mathrm{d}a\,\mathrm{d}b\,p(a,b|\mathrm{A,B})\,a\,b=\int \mathrm{d}\lambda\,p_{\lambda} \langle \mathrm{A}\rangle_{\lambda}\langle \mathrm{B}\rangle_{\lambda},
\ee
where we have introduced $\langle \mathrm{A}\rangle_{\lambda}\equiv \int\mathrm{d}a\,p(a|\mathrm{A},\lambda)\,a$ with similar notation for $\mathrm{B}$. Now consider the quantity 
\be\label{B}
\mathcal{B}\equiv| \langle \mathrm{A\otimes B}\rangle+\langle \mathrm{A'\otimes B}\rangle+\langle \mathrm{A\otimes B'}\rangle-\langle \mathrm{A'\otimes B'}\rangle|
\ee 
for observables $\{\mathrm{A,A',B,B'}\}$. Using Eq.~\eqref{<AB>} we can write $\mathcal{B}=\left|\int\mathrm{d}\lambda p_{\lambda}\mathcal{B}_{\lambda}\right|$, where 
\be\label{Blambda}
\mathcal{B}_{\lambda}=\langle \mathrm{A}\rangle_{\lambda}\left(\langle \mathrm{B}\rangle_{\lambda}+\langle \mathrm{B'}\rangle_{\lambda} \right)+\langle \mathrm{A'}\rangle_{\lambda}\left(\langle \mathrm{B}\rangle_{\lambda}-\langle \mathrm{B'}\rangle_{\lambda} \right).
\ee
By triangle inequality one has that $\mathcal{B}\leqslant \int \mathrm{d}\lambda p_{\lambda}|\mathcal{B}_{\lambda}|$ and
\be\label{|Blambda|}
|\mathcal{B}_{\lambda}|\leqslant \left|\langle \mathrm{A}\rangle_{\lambda}\right|\left|\langle \mathrm{B}\rangle_{\lambda}+\langle \mathrm{B'}\rangle_{\lambda} \right|+\left|\langle \mathrm{A'}\rangle_{\lambda}\right|\left|\langle \mathrm{B}\rangle_{\lambda}-\langle \mathrm{B'}\rangle_{\lambda} \right|.
\ee
We now examine the role of the spectrum. Let $o$ denotes the possible outcomes of a physical quantity $\mathrm{O}$. We assume the spectrum to be bounded as $o_{\min}\leqslant o\leqslant o_{\max}$ but leave open the possibility of it to be either discrete or continuous (we keep using notation for the latter case). For $\langle \mathrm{O}\rangle_{\lambda}=\int\mathrm{d}o\,p(o|\mathrm{O},\lambda)\,o$ it then follows that
\be\begin{array}{l} 
\langle \mathrm{O}\rangle_{\lambda}\leqslant \int\mathrm{d}o\,p(o|\mathrm{O},\lambda)\,o_{\max}=o_{\max},\\
\langle \mathrm{O}\rangle_{\lambda}\geqslant \int\mathrm{d}o\,p(o|\mathrm{O},\lambda)\,o_{\min}=o_{\min}.
\end{array}\ee 
This implies that $|\langle \mathrm{O}\rangle_{\lambda}|\leqslant \bar{o}$, where $\bar{o}\equiv\max\{|o_{\min}|,|o_{\max}| \}$. Let $\{\bar{a},\bar{a}',\bar{b},\bar{b}'\}$ be the respective bounds for the observables $\{\mathrm{A,A',B,B'}\}$. Inequality \eqref{|Blambda|} then becomes
\be\label{BmaxAA'}
|\mathcal{B}_{\lambda}|\leqslant \max\{\bar{a},\bar{a}'\}
\Big[\left|\langle \mathrm{B}\rangle_{\lambda}+\langle \mathrm{B'}\rangle_{\lambda} \right|+\left|\langle \mathrm{B}\rangle_{\lambda}-\langle \mathrm{B'}\rangle_{\lambda} \right| \Big].
\ee
The term in brackets can be treated as follows. Take $\alpha\equiv |u+v|+|u-v|\geqslant 0$ for $\{u,v\}\in\mathbbm{R}$. Simple algebra gives $\alpha^2=2(u^2+v^2+|u^2-v^2|)=4\max{\{u^2,v^2\}}$, which implies that $\alpha=2\max{\{|u|,|v|\}}$. Therefore, inequality \eqref{BmaxAA'} reduces to $|\mathcal{B}_{\lambda}|\leqslant 2\max{\{\bar{a},\bar{a}'\}}\max{\{\bar{b},\bar{b}'\}}$. By $\mathcal{B}\leqslant \int \mathrm{d}\lambda p_{\lambda}|\mathcal{B}_{\lambda}|$, we arrive at the final result:
\be\label{CHSHi}
\mathcal{B}\leqslant 2\max{\{\bar{a},\bar{a}'\}}\max{\{\bar{b},\bar{b}'\}}.
\ee
Our derivation intends to put in evidence two aspects that are not properly pointed out in the usual discussions of the CHSH inequality. First, inequality \eqref{CHSHi} contemplates observables with different spectra. Second, and most important, it applies to any bounded observables, regardless of whether their spectra are discrete or continuous. Therefore, instead of the spectral discreteness that appears in the vast majority of Bell tests realized to date, the key point behind the CHSH inequality \eqref{CHSHi} is {\em boundedness}. This observation is of particular relevance here because it will legitimize, in Sec.~\ref{CHVM}, submitting our CHVM to the CHSH test.

\section{Entanglement and nonlocality}
\label{QM}

In this section we introduce the measures of correlations we are interested in and the interaction model we are going to employ to study the dynamics of correlations. As far as pure states are concerned, the relevant quantum correlation is entanglement. Consider two partitions, $A$ and $B$, of a pure state $\ket{\psi_t}$ defined at a given instant $t$. The amount of entanglement $E(\ket{\psi_t})$ in $\ket{\psi_t}$ can be quantified by the von Neumann entropy of the reduced state $\rho_{A(B)}(t)=\text{Tr}_{B(A)}\ket{\psi_t}\bra{\psi_t}$, i.e., $E(\ket{\psi_t})=-\text{Tr}[\rho_s(t)\ln{\rho_s(t)}]$ for $s=\{A,B\}$. A very convenient alternative is to compute the entanglement $E$ of $\ket{\psi_t}$ by use of the linear entropy, i.e.,
\be 
E(\ket{\psi_t})=1-P_{q}[\rho_s(t)], 
\ee 
with $P_{q}(\rho)\equiv\text{Tr}(\rho^2)$ being the quantum purity of state $\rho$. 

For future convenience, especially in the context of the classical model to be developed, we now rephrase this measure through a different strategy. Take a generic Hamiltonian $\mathrm{H}=\mathrm{H}_A\otimes\mathbbm{1}_B+\mathbbm{1}_A\otimes \mathrm{H}_B+\mathrm{H}_{int}$, with which we can compute $\rho(t)$ from $\rho_0\in\mathcal{E}_A\otimes\mathcal{E}_B$. Now consider the following reduced density operator:
\be \label{tilderho}
\tilde{\rho}_{A(B)}(t)\equiv \text{Tr}_{B(A)}\left[ e^{-i(\mathrm{H}-\mathrm{H}_{int})t/\hbar}\rho_0e^{i(\mathrm{H}-\mathrm{H}_{int})t/\hbar}\right].
\ee 
This is just the reduced operator of the subsystem $s$ freely evolved in time, i.e., with no influence of coupling. It follows that the quantity $\text{Tr}_s\left[\tilde{\rho}_s^2(t)-\rho_s^2(t) \right]=P_{q}[\tilde{\rho}_s(t)]-P_{q}[\rho_s(t)]$ will signalize the influence of the interaction on the subsystem dynamics. Similar information will be retrieved from the normalized form of this quantity:
\be \label{Cs}
\mathcal{C}_s(t)\equiv 1-\frac{P_{q}[\rho_s(t)]}{P_{q}[\tilde{\rho}_s(t)]}.
\ee 
Issues reported in Ref.~\cite{angelo04} concerning asymmetries in classical models motivate us to work instead with the more democratic form
\be \label{Cq}
\mathcal{C}_{q}(t)\equiv \tfrac{1}{2}\big[\mathcal{C}_A(t)+\mathcal{C}_B(t)\big].
\ee 
The relevance of this quantifier is notorious in the context of initially-separable pure states. In this case, and because of unitary evolution, one has that $P_{q}[\tilde{\rho}_s(t)]=1$ and, therefore, $\mathcal{C}_{q}(t)=\mathcal{C}_s(t)=E[\rho(t)]$. In other words, $\mathcal{C}_{q}(t)$ is nothing but an entanglement quantifier for dynamics of initially-separable pure states. As mentioned before, however, the elaborate form \eqref{Cq} intends to give a better chance to the classical counterpart that will be developed soon.

\begin{figure}[htb]
\includegraphics[width=\columnwidth]{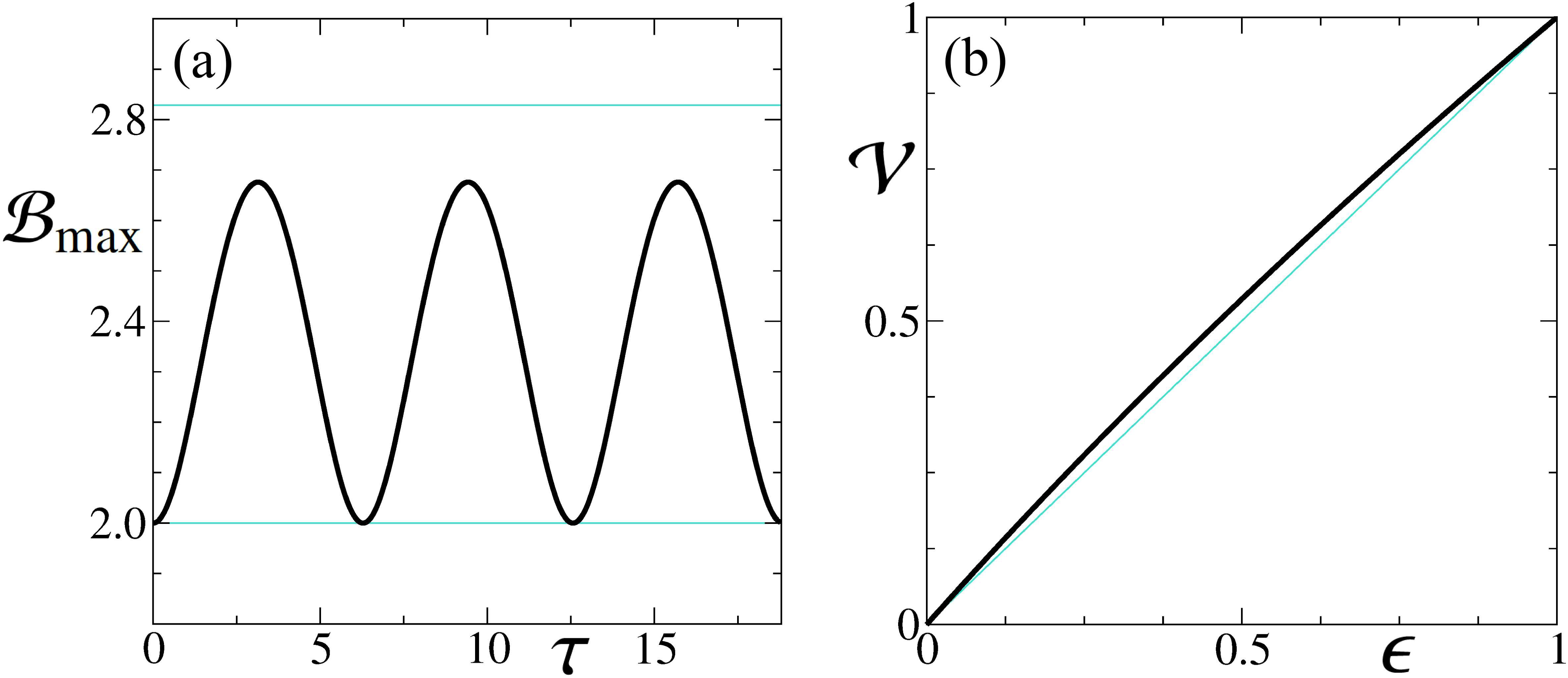}
\caption{(Color online) (a) $\mathcal{B}_{\max}$ [Eq.~\eqref{Bmax}] as a function of the dimensionless time $\tau=\xi t$ (thick black line). Except for $\tau=2\pi n$ ($n\in\mathbb{N}$), nonlocality is always present in the dynamics induced by the coupling \eqref{Hsigmaz}. Turquoise thin lines indicate the minimum value assumed by $\mathcal{B}_{\max}$ during the dynamics and Tsirelson's bound $(2\sqrt{2})$. (b) Parametric plot for the violation quantifier $\mathcal{V}$ [see Eq.~\eqref{calV}] and the normalized entanglement $\epsilon=\sin^2\left(\tau/2\right)$ (thick black line). The turquoise thin line, which represents $\mathcal{V}=\epsilon$, is used just for appreciation of the result. In these simulations, $w_{0s}=1/\sqrt{2}$ and $\tau\in[0,6\pi]$.}
\label{fig1}
\end{figure}

As far as the physical coupling is concerned, the chosen scenario is the simplest and as well established as possible. Consider two spin-1/2 systems, designated as $A$ and $B$, initially prepared in a separable pure state
\begin{subequations}
\beq 
\ket{\psi_0}&=&\ket{w_{0A}}\otimes\ket{w_{0B}},\label{psi0} \\ \nonumber \\
\ket{w_{0s}}&=&\frac{\ket{\tfrac{1}{2},-\tfrac{1}{2}}+w_{0s}\ket{\tfrac{1}{2},+\tfrac{1}{2}}}{\sqrt{1+|w_{0s}|^2}},\label{|w0s>}
\eeq 
\end{subequations}
with $w_{0s}\in\mathbb{C}$ and $\mathrm{J}_\mathrm{z}^s\ket{\tfrac{1}{2},\pm\tfrac{1}{2}}=\pm\tfrac{\hbar}{2}\ket{\tfrac{1}{2},\pm\tfrac{1}{2}}$. It is sufficient for our purposes to take the following coupling:
\be \label{Hsigmaz}
\mathrm{H}_{int}=\frac{\xi}{\hbar}\mathrm{J}^A_{\mathrm{z}}\otimes\mathrm{J}^B_{\mathrm{z}}=\tfrac{\hbar\xi}{4}\,\left(\hat{\mathrm{z}}\cdot\vec{\sigma}^A\right)\otimes\left(\hat{\mathrm{z}}\cdot\vec{\sigma}^B\right),
\ee 
for some interaction strength $\xi\in\mathbb{R}$. Throughout this work, $\vec{\sigma}^s=(\sigma_{\mathrm{x}}^s,\sigma_{\mathrm{y}}^s,\sigma_{\mathrm{z}}^s)$ denotes the vector whose components are Pauli's matrices acting on the Hilbert space $\mathcal{E}_s$ of the subsystem $s$ and $\{\hat{\mathrm{x}},\hat{\mathrm{y}},\hat{\mathrm{z}}\}$ is the orthonormal basis employed to span vectors in $\mathbb{R}^3$. For simplicity, free dynamics is neglected. Given the above, the calculation of $\ket{\psi_t}=\mathrm{U}(t)\ket{\psi_0}$ with $\mathrm{U}(t)=e^{-i\mathrm{H}t/\hbar}$ and $H=H_{int}$ is a simple exercise. The calculation of the quantum correlations \eqref{Cq} is also straightforward and leads to
\be \label{Cqresult}
\mathcal{C}_{q}(\tau)=\frac{8|w_{0A}|^2|w_{0B}|^2}{(1+|w_{0A}|^2)^2(1+|w_{0B}|^2)^2}\sin^2\left(\tfrac{\tau}{2}\right),
\ee 
with $\tau=\xi t$ being a dimensionless time scale. By Gisin's theorem one has that whenever $\mathcal{C}_{q}>0$, the CHSH must be violated by some measurement setting. Therefore, nonlocality is expected for all times except for $\tau=2\pi n$ $(n\in\mathbb{N})$. Let us prove this explicitly for $w_{0A}=w_{0B}=1/\sqrt{2}$. In reference to the CHSH inequality \eqref{CHSHi}, we consider the dimensionless observables 
\be \label{AA'BB'} \begin{array}{lll}
\mathrm{A}=\hat{n}(\theta_a,\phi_a)\cdot\vec{\sigma}^A, & \qquad & \mathrm{A}'=\hat{n}(\theta_{a'},\phi_{a'})\cdot\vec{\sigma}^A, \\
\mathrm{B}=\hat{n}(\theta_b,\phi_b)\cdot\vec{\sigma}^B, & \qquad & \mathrm{B}'=\hat{n}(\theta_{b'},\phi_{b'})\cdot\vec{\sigma}^B,
\end{array} \ee 
where $\hat{n}(\theta,\phi)=(\sin{\theta}\cos{\phi},\sin{\theta}\sin{\phi},\cos{\theta})$. Without loss of generality, we choose the coordinate system in a way such that $\hat{n}(\theta_a,\phi_a)=\hat{\mathrm{z}}$. We then optimize the quantity \eqref{B} over the measurement setting $\mu=\{\theta_{a'},\theta_b,\theta_{b'},\phi_{a'},\phi_b,\phi_{b'} \}$ so as to obtain
\be \label{Bmax}
\mathcal{B}_{\max}(\tau)=\max_{\mu}\mathcal{B}(\tau),
\ee 
where optimization is to be conducted for each instant of time. This constitutes a $\mu$-independent quantity, which allows us to infer the presence of nonlocality regardless of the measurement setting. In fact, $\mathcal{B}_{\max}>2$ emerges as a sufficient condition for the presence of nonlocality, for this inequality guarantees that there exists at least one measurement setting for which the CHSH is violated. Figure~\ref{fig1} presents numerical results for $\mathcal{B}_{\max}$ as a function of $\tau$ [Fig.~\ref{fig1}(a)], showing that CHSH-inequality violations occur for all times except at $\tau=2\pi n$, just like entanglement, and a direct comparison [Fig.~\ref{fig1}(b)] between the normalized entanglement $\epsilon(\tau)\equiv \mathcal{C}_{q}(\tau)/\max_{\tau}\mathcal{C}_{q}(\tau)=\sin^2\left(\tau/2\right)$ and the normalized quantifier of CHSH-inequality violation
\be \label{calV}
\mathcal{V}(\tau)\equiv \frac{\mathcal{B}_{\max}(\tau)-2}{\max_{\tau}\mathcal{B}_{\max}(\tau)-2}.
\ee
The results make it evident that $\mathcal{V}(\tau)$ can be written as a monotonically increasing function of $\epsilon(\tau)$, which clearly demonstrates the equivalence between entanglement and nonlocality, as anticipated by Gisin's theorem.

\section{Classical-hidden-variable model}
\label{CHVM}

In this section, we construct our CHVM for the dynamics of two spins $j$. It should be stated from the outset that our model aims at furnishing predictions only for expectations values and correlations; that is, it is not intended to provide the individual outcomes of measurements. Also, it is inspired by Ehrenfest's theorem, which guarantees that classical trajectories can mimic expectations values well under certain conditions. 

A key ingredient here is the spin coherent state~\cite{radcliffe71,gazeau09}
\be\label{|w>}
\ket{w}=\frac{e^{w \mathrm{J}_+/\hbar}}{\mathcal{N}_j} \ket{j,-j}
=\frac{1}{\mathcal{N}_j}\sum_{n=0}^{2j}w^n\left(\begin{matrix} 2j \\ n \end{matrix}\right)^{\!\!\!1/2}\ket{j,-j+n},
\ee
where $\mathrm{J}_{\pm}\equiv \mathrm{J}_{\mathrm{x}}+i\mathrm{J}_{\mathrm{y}}$ are the usual ladder operators, $\{\ket{j,m}\}$ is the simultaneous eigenbasis of $\{\mathrm{J}^2,\mathrm{J}_{\mathrm{z}}\}$ with respective eigenvalues $\{j(j+1)\hbar^2,m\hbar\}$, $\mathcal{N}_j=(1+|w|^2)^j$ is a normalization term, and $w$ is a complex number. Spin coherent states are non-orthogonal and form the completeness relation
\be \label{1}
\frac{(2j+1)}{\pi}\int\frac{\mathrm{d}\text{Re}{(w)}\,\mathrm{d}\text{Im}{(w)}}{(1+|w|^2)^2}\ket{w}\bra{w}=\mathbbm{1}.
\ee 
For the coherent state \eqref{|w>} one shows by direct calculation that
\begin{subequations}\label{expectations}
\beq
&&\langle\vec{\mathrm{J}}\,\rangle=\hbar j\left(\frac{w+w^*}{1+|w|^2},\frac{i(w-w^*)}{1+|w|^2},\frac{|w|^2-1}{1+|w|^2}\right),\\
&&(\Delta \vec{\mathrm{J}}\,)^2\equiv\langle \vec{\mathrm{J}}\cdot\vec{\mathrm{J}}\,\rangle-\langle \vec{\mathrm{J}}\,\rangle\cdot\langle \vec{\mathrm{J}}\,\rangle=\sum\limits_{\text{\tiny $r=\mathrm{x,y,z}$}}(\Delta \mathrm{J}_r)^2=\hbar^2j,\label{DJ}
\eeq
\end{subequations}
where $\vec{\mathrm{J}}=(\mathrm{J}_{\mathrm{x}},\mathrm{J}_{\mathrm{y}},\mathrm{J}_{\mathrm{z}})$ and $(\Delta \mathrm{J}_r)^2=\langle \mathrm{J}_r^2\rangle-\langle \mathrm{J}_r\rangle^2$. Classical variables are introduced in our approach through a proper conjugation of the limits $\hbar\to 0$ and $j\to 1/\hbar$ along with the following transformations. First, $w=e^{-i\phi}\cot{\tfrac{\theta}{2}}$ parametrizes the Riemann sphere through a stereographic projection of unit vectors $\hat{n}(\theta,\phi)$ onto $\mathbb{C}$ with respect to the north pole. Second, we introduce canonical coordinates $\vec{x}\equiv(q,p)\equiv(\phi,\cos{\theta})$, with $q\in [0,2\pi]$ and $p\in [-1,1]$. Thus, for a generic function $F(\mathrm{J}_r)$ of spin observables $\mathrm{J}_r$, the corresponding classical function $\mathcal{F}$ is given by
\be\label{calF}
\mathcal{F}(\vec{x})\equiv\lim_{\hbar\to 0}\lim_{j\to 1/\hbar}\Big[F\big(\bra{w}\mathrm{J}_r\ket{w}\big)\Big]_{w=e^{-iq}\sqrt{\tfrac{1+p}{1-p}}}.
\ee 
That this prescription produces the desirable effect is immediately seen from its application to Eqs.~\eqref{expectations}. We arrive at the ``classical spin'' $\vec{\mathcal{J}}=\lim_{\hbar\to 0}\lim_{j\to 1/\hbar}\langle\vec{\mathrm{J}}\,\rangle=\hat{n}(\theta,\phi)$, which can be written as
\be\label{calJ}
\displaystyle\vec{\mathcal{J}}(\vec{x})=(\sqrt{1-p^2}\cos{q},\sqrt{1-p^2}\sin{q},p).
\ee 
and $(\Delta \vec{\mathcal{J}})^2=\lim_{\hbar\to 0}\lim_{j\to 1/\hbar}(\Delta \vec{\mathrm{J}})^2=0$. Clearly, $\vec{\mathcal{J}}$ is a vector with norm $j\hbar=1$ and null variance. This demands, via Eq.~\eqref{DJ}, that $\Delta \mathcal{J}_{\mathrm{x,y,z}}=0$, in clear violation of Heisenberg's principle. This shows the strict sense in which the spin vector $\vec{\mathcal{J}}=(\mathcal{J}_{\mathrm{x}},\mathcal{J}_{\mathrm{y}},\mathcal{J}_{\mathrm{z}})$ is classical: its components all assume well-defined values simultaneously.

Concerning the physical Hamiltonian and its implied dynamics, we restrict our analysis to two-spin couplings as
\be \label{H}
\mathrm{H}=\tfrac{\xi}{\hbar}\,\vec{\mathrm{J}}^A\cdot\vec{\mathrm{J}}^B=\tfrac{\xi}{\hbar} \sum_{r=\mathrm{x,y,z}}\mathrm{J}_r^A\otimes\mathrm{J}_r^B,
\ee 
where $\xi$ is the interaction strength and $\mathrm{J}_r^{A,B}$ are spin components for subsystems $A$ and $B$. The expectation values in $\ket{w}$ of Heisenberg's equations for this Hamiltonian read
\be\label{Heisenberg}
\langle\dot{\vec{\mathrm{J}}}^{A(B)}\rangle=\xi\langle \vec{\mathrm{J}}^{B(A)}\times\vec{\mathrm{J}}^{A(B)}\rangle.
\ee 
Notation is such that $[\vec{\mathrm{J}}^{A}\times\vec{\mathrm{J}}^{B}]_{\mathrm{z}}=\mathrm{J}^A_{\mathrm{x}}\otimes\mathrm{J}^B_{\mathrm{y}}-\mathrm{J}^A_{\mathrm{y}}\otimes\mathrm{J}^B_{\mathrm{x}}$, with similar expressions for the other components. The corresponding classical dynamics emerges by the prescription \eqref{calF}. To synchronize the classical time scale with the quantum one, we construct the classical Hamiltonian function via 
\be \label{classicalH}
\mathcal{H}=\lim_{\hbar\to 0}\lim_{j\to1/\hbar}\bra{w}\,\tfrac{\mathrm{H}}{j}\,\ket{w}.
\ee 
We obtain that $\mathcal{H}=\xi\hat{n}^A\cdot\hat{n}^B$  and $\dot{\hat{n}}^{A(B)}=\xi \hat{n}^{B(A)}\times\hat{n}^{A(B)}$, which in terms of the phase-space variables $\vec{x}_A=(q_A,p_A)$ and $\vec{x}_B=(q_B,p_B)$ assume the usual Hamiltonian structure
\begin{subequations}\label{Hamilton}
\beq
&&\mathcal{H}(\vec{x}_A,\vec{x}_B)=\xi\vec{\mathcal{J}}(\vec{x}_A)\cdot\vec{\mathcal{J}}(\vec{x}_B), \\
&&\dot{\vec{x}}_s=\mathfrak{I}\,\nabla_s\mathcal{H}(\vec{x}_A,\vec{x}_B),\label{Heqs}
\eeq 
\end{subequations}
where
\be 
\vec{x}_s=\left(\begin{matrix} q_s \\ p_s\end{matrix}\right),\quad \mathfrak{I}=\left(\begin{matrix} 0 & 1 \\ -1 & 0\end{matrix}\right),\quad \nabla_s=\left(\begin{matrix} \partial_{q_s} \\ \partial_{p_s}\end{matrix}\right),
\ee
and $s=\{A,B\}$. $\mathfrak{I}$ is the usual $2\times 2$ symplectic matrix. For convenience, throughout this paper we use $\vec{x}_s$ to interchangeably denote either the vector $(q_s,p_s)$ or the column matrix $(q_s \,\,\,\, p_s)^T$ ($T$ stands for transposition).

In the same spirit as Bohmian's interpretation, here the Hamiltonian trajectories $\vec{x}=(\vec{x}_A,\vec{x}_B)$ play the role of {\em hidden variables} (``hidden'' in the sense of not being available in quantum theory). Also, one may readily verify from Hamilton's equations~\eqref{Heqs} that these trajectories are, in general, {\em nonlocal}. Take, e.g., 
\be\label{gen_force}
\dot{p}_A=-\partial_{q_A}\mathcal{H}=-\xi\left[\partial_{q_A}\vec{\mathcal{J}}(\vec{x}_A)\right]\cdot\vec{\mathcal{J}}(\vec{x}_B).
\ee 
As for Bohmian trajectories, here we see that the generalized force $\dot{p}_A$ acting on spin $A$ depends on the generalized position $q_B$ of spin $B$, so that any change in $q_B$ instantaneously affects the spin $A$ even when they are arbitrarily far apart. That is, the interaction model \eqref{H} as it stands is nonlocal. Since we are interested in constructing a CHVM that is manifestly {\em local}, we further restrict our interaction model to
\be \label{HcalH}
\mathrm{H}=\tfrac{\xi}{\hbar}\,\left(\hat{\mathrm{z}}\cdot\vec{\mathrm{J}}^A\right)\otimes\left(\hat{\mathrm{z}}\cdot\vec{\mathrm{J}}^B\right) \quad \leftrightarrow \quad \mathcal{H}=\xi p_Ap_B.
\ee 
In this case, $\dot{p}_{A,B}=0$, and the aforementioned nonlocality problem disappears. Also, for simplicity, hereafter we focus on two-qubit systems, i.e., $j^s=1/2$ and $\mathrm{J}_{\mathrm{z}}^s=\hbar\sigma_{\mathrm{z}}^s/2$.

Having introduced the local hidden variables of the model, we now implement statistics. To this end, we employ the Liouville theory, which guarantees that any probability distribution $\wp_0(\vec{x})$ can be evolved in time via Liouville's equation, $\partial_t\wp=\{\mathcal{H},\wp\}$, in agreement with the Hamiltonian flow $\Phi_t$ which propagates any phase-space point as $\vec{x} \mapsto \vec{x}(t)=\Phi_t(\vec{x})$. The formal solution to Liouville's equation emerges as follows. In terms of Poisson's brackets, the Hamilton equations read $\dot{\vec{x}}=-\mathcal{L}\vec{x}$, with the Liouville operator $\mathcal{L}\equiv\{\mathcal{H},\bullet\}=\sum_{s}\left(\partial_{q_s}\mathcal{H}\partial_{p_s}\bullet-\partial_{p_s}\mathcal{H}\partial_{q_s}\bullet\right)$. It follows that $\vec{x}(t)=e^{-\mathcal{L}t}\vec{x}$. Liouville's equation can now be written as $\partial_t\wp=\mathcal{L}\wp$, whose formal solution is given by
\be \label{wp(t)}
\wp(\vec{x},t)=e^{\mathcal{L}t}\wp(\vec{x},0)=\wp_0(\vec{x}(-t)),
\ee 
where $\wp(\vec{x},0)=\wp_0(\vec{x})$. The initial probability distribution $\wp_0(\vec{x})$, which in our model is to be viewed as an epistemic state deriving from subjective ignorance of the (ontic) hidden state $\vec{x}$ of the system, is constructed by inspection of two natural candidates, namely, the Husimi and Wigner~\cite{dowling94} functions associated with the initial quantum state \eqref{psi0} and the parametrization $w_{0s}=e^{-iq_{0s}}\sqrt{(1+p_{0s})/(1-p_{0s})}$:
\begin{subequations}\label{HW}
\beq 
H(\vec{x}_s,\vec{x}_{0s})&=&\frac{1+f(\vec{x}_s,\vec{x}_{0s})}{4\pi},\label{Husimi} \\
W(\vec{x}_s,\vec{x}_{0s})&=&\frac{1+\sqrt{3}f(\vec{x}_s,\vec{x}_{0s})}{4\pi},\label{W}
\eeq 
\end{subequations}
where
\be 
f(\vec{x}_s,\vec{x}_{0s})=p_sp_{0s}+\cos{(q_s-q_{0s})}\sqrt{(1-p_s^2)(1-p_{0s}^2)}.
\ee 
Inspired by the obvious similarity between the functions, we propose to take as the initial probability distribution
\begin{subequations}
\beq
&&\wp_0(\vec{x})=\wp_{\delta}(\vec{x}_A,\vec{x}_{0A})\wp_{\delta}(\vec{x}_B,\vec{x}_{0B}),\label{wp0} \\
&&\wp_{\delta}(\vec{x}_s,\vec{x}_{0s})=\frac{1+\delta f(\vec{x}_s,\vec{x}_{0s})}{4\pi}.\label{wpd}
\eeq 
\end{subequations}
This choice is interesting because it introduces an extra hidden variable $(\delta)$ to our CHVM. The marginal distributions $\wp_{\delta}$ will be non-negative whenever $1+\delta \min_{\vec{x},\vec{x}_0} f\geqslant 0$. Because $\min_{\vec{x},\vec{x}_0} f=-\max_{\vec{x},\vec{x}_{0s}}|f|=-1$, one has that $\delta\in[0,1]$. This reveals that the Wigner function \eqref{W} is not a suitable candidate for classical probability distribution after all.

To compute the time-evolved classical distribution via formula \eqref{wp(t)}, we need to determine the hidden trajectory $\vec{x}(t)$. Calculating Hamilton's equations for the Hamiltonian \eqref{HcalH}, we obtain
\be
\vec{x}(t)=\left(\begin{matrix} q_A(t) \\ p_A(t) \\ q_B(t) \\ p_B(t)\end{matrix}\right)=\left(\begin{matrix} 1 & 0 & 0 & \xi t \\ 0 & 1 & 0 & 0 \\ 0 & \xi t & 1 & 0 \\ 0 & 0 & 0 & 1\end{matrix}\right)\left(\begin{matrix} q_A\\ p_A \\ q_B \\ p_B\end{matrix}\right)\equiv M(t)\,\vec{x}.
\ee 
With that, we finish the construction of our CHVM. Now we have at hand all the tools to compute expectation values at any times and, therefore, to submit our CHVM to the CHSH test. Before doing so, however, it is convenient to calculate the quantum counterpart of the correlation measure \eqref{Cq}. To this end, we first introduce, guided by the results reported in Ref.~\cite{angelo04}, the {\em classical purity}
\be \label{Pcl}
P_{\text{cl}}[\wp(\vec{x},t)]\equiv \frac{\int \mathrm{d}^4\vec{x}\,\wp^2(\vec{x},t)}{\int \mathrm{d}^4\vec{x}\,\wp^2(\vec{x},0)}.
\ee 
In the domain of classical probability distributions, there is no analog for the idempotency property $\rho^2=\rho$ of pure states, so one could not imitate the maximum quantum purity $P_{q}(\rho)=1$ within this formalism without the normalization term appearing above. Clearly, this normalization guarantees, by construction, a fair comparison between the classical and quantum purities at least at $t=0$, the instant at which $P_{\text{cl}}[\wp(\vec{x},0)]=1$. Of course, this classical mimic makes sense only when the corresponding quantum scenario involves pure states. Following the definitions \eqref{tilderho}-\eqref{Cq}, we introduce
\be 
\tilde{\wp}_{A(B)}(\vec{x}_{A(B)},t)=\int\mathrm{d}^2\vec{x}_{B(A)}\lim_{\mathcal{H}_{int}\to 0}\wp(\vec{x},t),
\ee 
where $\wp(\vec{x},t)=\wp_0(\vec{x}(-t))$ is the joint probability distribution evolved in time with the Hamiltonian $\mathcal{H}=\mathcal{H}_A+\mathcal{H}_B+\mathcal{H}_{int}$. To capture the influence of the interaction in the dynamics of the subsystem $s$ we take the quantity
\be 
\mathcal{C}_s^{\text{cl}}(t)\equiv1-\frac{P_{\text{cl}}[\wp_s(\vec{x}_s,t)]}{P_{\text{cl}}[\tilde{\wp}_s(\vec{x}_s,t)]}.
\ee 
Finally, in full analogy with the quantum-correlation quantifier \eqref{Cq}, we introduce the measure
\be \label{Ccl}
\mathcal{C}_{\mathrm{cl}}(t)\equiv\tfrac{1}{2}\left[\mathcal{C}_A^{cl}(t)+\mathcal{C}_B^{cl}(t) \right].
\ee 
It is instructive to notice that this is a measure of inseparability of the classical probability distribution. Assume that $\mathcal{H}_{int}=0$ and $\wp_0(\vec{x})=\wp_{0A}(\vec{x}_A)\wp_{0B}(\vec{x}_B)$. Because there is no coupling, one has that $\wp(\vec{x},t)=\wp_{0A}(\vec{x}_A(-t))\wp_{0B}(\vec{x}_B(-t))$. It follows that $\tilde{\wp}_s(\vec{x}_s,t)=\wp_s(\vec{x}_s,t)$, which yield $\mathcal{C}_s^{\text{cl}}(t)=0$ and, therefore, $\mathcal{C}_{\text{cl}}(t)=0$. For non-null couplings, we will generally have $\mathcal{C}_{\text{cl}}(t)>0$, as we will see next (see Refs.~\cite{angelo04,angelo05,matzkin11,casati12} for numerical illustrations with similar quantities). 

\begin{figure}[htb]
\includegraphics[width=\columnwidth]{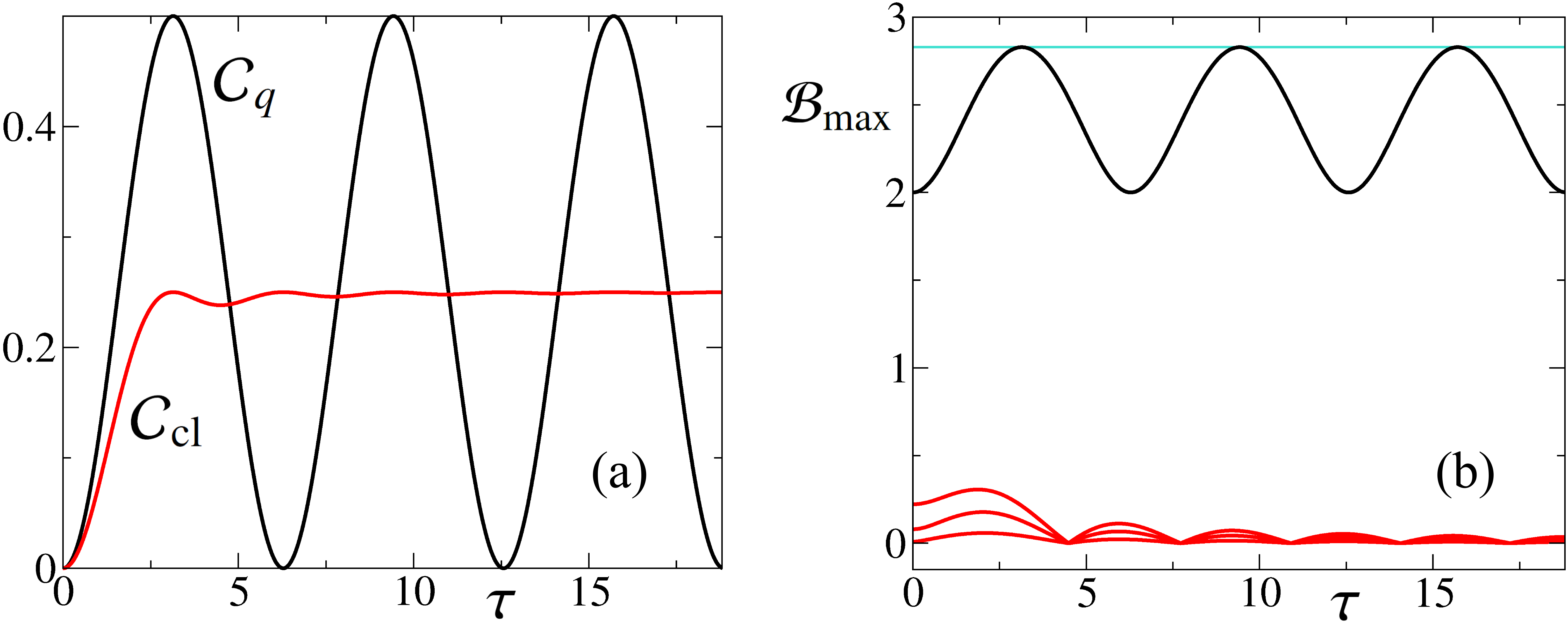}
\caption{(Color online) (a) Entanglement $\mathcal{C}_{q}(\tau)$ (black periodic line) and classical inseparability $\mathcal{C}_{\text{cl}}(\tau)$ (red line) as a function of the dimensionless time $\tau=\xi t$.  (b) $\mathcal{B}_{\max}(\tau)$ for the quantum operators~\eqref{AA'BB'} (black solid line) and for the classical observables \eqref{AA'BB'cl} (red lines) and classical distributions \eqref{wp0} with $\delta=0.2$ (lower red line), $\delta=0.6$ (middle red line), and $\delta=1.0$ (upper red line), as a function of $\tau$.  The Tsirelson bound $(2\sqrt{2})$ is also shown (turquoise solid line). In these simulations, $w_{0s}=1$, $q_{0s}=p_{0s}=0$, and $\tau\in[0,6\pi]$.}
\label{fig2}
\end{figure}

We now present the results obtained with our CHVM model. The resulting formula for the classical inseparability measure can be written as
\begin{subequations}
\be \label{Cclresult}
\mathcal{C}_{\text{cl}}(\tau)=\frac{\delta^2}{3+\delta^2}\left[\mathcal{X}(\tau)+\frac{\delta^2}{\tau^2}\mathcal{Y}(\tau) \right],
\ee
where
\beq 
&&\mathcal{X}(\tau)=(1-\alpha)\left[1-\left(\tfrac{\sin{\tau}}{\tau}\right)^2 \right],\\
&&\mathcal{Y}(\tau)=(\beta-\alpha)\left(\cos{\tau}-\tfrac{\sin{\tau}}{\tau} \right)^2,
\eeq
\end{subequations}
$\alpha=\tfrac{1}{2}(p_{0A}^2+p_{0B}^2)$, and $\beta=p_{0A}^2p_{0B}^2$. Notice that $\mathcal{X}(0)=\mathcal{Y}(0)=0$ and $\{\alpha,\beta,\mathcal{X}\}\in [0,1]$. The above result should be compared with its quantum counterpart \eqref{Cqresult}, which in terms of the parametrization $w=e^{-iq}\sqrt{(1+p)/(1-p)}$ reduces to
\be 
\mathcal{C}_{q}(\tau)=\tfrac{1}{2}\left(1-p_{0A}^2\right)\left(1-p_{0B}^2\right)\sin^2\left(\tfrac{\tau}{2}\right).
\ee 
Likewise the quantum result, $\mathcal{C}_{\text{cl}}(\tau)$ does not depend on $\{q_{0A},q_{0B}\}$, which reflects symmetry around the $\mathrm{z}$ axis. Also, it is fully symmetrical under permutations $p_{0A}\leftrightarrow p_{0B}$ of the subsystem, which is desirable for a measure that is intended to mimic entanglement. This shows that the symmetrization scheme proposed in Eq.~\eqref{Ccl} is advantageous in relation to previous approaches~\cite{angelo04}. There are, however, many aspects with respect to which the classical measure significantly departures from the quantum one. First, the classical solution is marked by an ``irreversible'' behavior, which derives from terms involving powers of $\tau$, in clear contrast to the periodic quantum result. In fact, the classical measure asymptotically converges to $\mathcal{C}_{\text{cl}}(\infty)=\delta^2(1-\alpha)/(3+\delta^2)$ and cannot reproduce the quantum recurrences $\mathcal{C}_{q}(2\pi n)=0$  ($\forall\,\{n,p_{0s}\}$), which never ceases to happen. Second, entanglement vanishes for either $|p_{0A}|=1$ {\em or} $|p_{0B}|=1$, whereas its classical counterpart vanishes only if $|p_{0A}|=1$ {\em and} $|p_{0B}|=1$. Third, the classical measure is manifestly dependent on the the extra hidden variable $\delta$, which regulates the initial distribution adopted. Fourth, the maximum value that entanglement can reach in this system is $1/2$, whereas for the classical measure the maximum is $1/4$ (for $\delta=1$). This suggests that a better chance can be given to the classical measure if the comparison is made in terms of normalized measures (as we show below). Figure \ref{fig2}(a) provides an illustration of the behavior of the measures $\mathcal{C}_{q,cl}(\tau)$ as a function of $\tau$.

For our purposes in this work, what is most interesting about the classical inseparability \eqref{Ccl} is that it is able to satisfactorily report on the dynamics of quantum entanglement at short times. One may show from the analytical results that, for $\tau\ll 1$, one has that $\mathcal{C}_{q}(\tau)=\omega_q \tau^2$ and $\mathcal{C}_{\text{cl}}(\tau)=\omega_{cl}\tau^2$, with $\omega_q=\tfrac{1}{8}(1-2\alpha+\beta)$ and $\omega_{cl}=\tfrac{1}{36}(3-4\alpha+\beta)$. This shows that the classical measure contains the fundamental information about the entanglement dynamics since they can be written as monotonically increasing functions of each other. In fact, we have numerically checked that there is a clear monotone correspondence between the normalized entanglement $\epsilon(\tau)=\sin^2(\tau/2)$ and the normalized classical inseparability $\mathfrak{c}(\tau)\equiv \mathcal{C}_{\text{cl}}(\tau)/\mathcal{N}$ in the domain $\tau\in [0,\pi]$, for all $\{p_{0s},\delta\}$. The normalization factor is $\mathcal{N}=\max_{\tau}\mathcal{C}_{\text{cl}}(\tau)=\mathcal{C}_{\text{cl}}(\infty)$.

This result sounds paradoxically in light of the following reasoning. In the domain $\tau\in[0,\pi]$, we have a CHVM based on local classical trajectories that is capable of correctly informing us about the entanglement dynamics. By virtue of Gisin's theorem [see Fig.~\ref{fig1}(b)], entanglement is equivalent to nonlocality. Therefore, we have a {\em local} CHVM that plays the role of a faithful informer of the {\em nonlocality} dynamics. Naturally, the question arises as to whether such a model would be at odds with Bell's theorem. We then submit our CHVM to the CHSH test. Notice that, as shown in Sec.~\ref{CHSH}, this test will indeed be meaningful as long as we look at bounded observables. To satisfy this condition, we propose, in direct analogy with the quantum scenario [see Eqs.~\eqref{AA'BB'}], to construct a CHSH test based on the observables
\be \label{AA'BB'cl} \begin{array}{lll}
\mathfrak{a}=\hat{n}(\theta_a,\phi_a)\cdot\vec{\mathcal{J}}(\vec{x}_A), &  & \mathfrak{a}'=\hat{n}(\theta_{a'},\phi_{a'})\cdot\vec{\mathcal{J}}(\vec{x}_A), \\
\mathfrak{b}=\hat{n}(\theta_b,\phi_b)\cdot\vec{\mathcal{J}}(\vec{x}_B), & & \mathfrak{b}'=\hat{n}(\theta_{b'},\phi_{b'})\cdot\vec{\mathcal{J}}(\vec{x}_B),
\end{array} \ee 
where we choose, without loss of generality, $\hat{n}(\theta_a,\phi_a)=\hat{\mathrm{z}}$. As required, one has that $\{\mathfrak{a,a',b,b'}\}\in [-1,1]$. Expectation values are here computed as
\be 
\langle\mathfrak{ab}\rangle(\tau)=\int\mathrm{d}^4\vec{x}\,\,\,\,\wp(\vec{x},\tau)\,\,\mathfrak{ab}.
\ee 
Nonlocality will be present if $\mathcal{B}_{\max}> 2$, with $\mathcal{B}_{\max}$ given as in Eq.~\eqref{Bmax} and $\mathcal{B}(\tau)=\langle \mathfrak{a(b+b')+a'(b-b')} \rangle$. We look at the case for which the entanglement production is enhanced, namely, $w_{0s}=1$, which corresponds to $q_{0s}=p_{0s}=0$. Numerical results for $\mathcal{B}_{\max}$ are presented in Fig.~\ref{fig2}(a) for both the quantum and the classical scenarios. Interestingly, we see that even though our CHVM can mimic of the entanglement dynamics in the short-time regime [see the domain $\tau\in[0,\pi]$ in Fig.~\ref{fig2}(a)] well, it is very far from violating the CHSH inequality [Fig.~\ref{fig2}(b)], regardless of the value taken for the hidden variable $\delta$. This result suggests that our CHVM is {\em local} not only at the level of the hidden trajectories $\vec{x}$ [as discussed below Eq.~\eqref{HcalH}], but also in the perspective of not violating the Bell-locality hypothesis \eqref{Blocality}.

As pointed out above, the classical results exhibit a sort of ``irreversible'' behavior, which is marked by a gradual damping of oscillatory regimes, in contrast to pronounced quantum recurrences. This discrepancy can be attributed to the absence, in the classical formalism, of {\em interference phenomena}, which are abundant in the quantum domain. To appreciate the point, let us focus on the dynamics of the reduced state $\rho_A(\tau)$. At $\tau=0$, it is a pure coherent state, $\ket{w_{0A}}$, whose phase-space representation is given, say, by the Husimi distribution $H(\vec{x}_A,\tau=0)$ [see Eq.~\eqref{Husimi}]. As time evolution takes place, $\rho(t)$ gets entangled, and the reduced state $\rho_A(\tau)$ is no longer pure. Its ``phase-space wave'' $H(\vec{x}_A,t)$ spreads out and loses oscillatory structures, which is an expression of the reduced-state decoherence. For the interaction model in question, the dynamics is periodic, i.e., $\rho(\tau)=\rho(\tau+4\pi n)$ $(\forall n\in\mathbb{Z})$. This means that $\rho(4\pi)=\rho(0)$; hence, $\rho_A(4\pi)$ has to be again the initial pure state $\ket{w_{0A}}$, with its corresponding coherent wave $H(\vec{x}_A,0)$. This reconstruction process, which can be interpreted as being supported by quantum interference, has no classical analog. This can be checked by direct comparison of the classical marginal distribution $\wp(\vec{x}_A,\tau)$ with 
\be
H(\vec{x}_A,\tau)=\left[\frac{\bra{w_A}\rho_A(\tau)\ket{w_A}}{4\pi}\right]_{w_A=e^{-iq_A}\sqrt{\frac{1+p_A}{1-p_A}}},
\ee 
the Husimi representation of the reduced state $\rho_A(\tau)$. The resulting distributions for arbitrary $\tau$ are lengthy and not insightful, so they will be omitted. For our discussion, it is sufficient to consider the instants $\tau_n=n\pi/2$ and $p_{0s}=0$, in which case one has that
\be\label{Hwp}
\left[\begin{matrix} H(\vec{x}_A,\tau_n) \\ \wp(\vec{x}_A,\tau_n)\end{matrix}\right]=\frac{1}{4\pi}+\left[\begin{matrix} \Gamma_{q}(n) \\ \Gamma_{\text{cl}}(n)\end{matrix}\right] \frac{\sqrt{1-p_A^2}\cos{(q_A-q_{0A})}}{4\pi},
\ee
where 
\be 
\Gamma_{q}(n)=\cos{\left(\frac{n\pi}{4}\right)}, \quad\quad \Gamma_{\text{cl}}(n)=\delta\frac{\sin{(n\pi/2)}}{n\pi/2}.
\ee 
Since $q_{0A}$ plays the role of a relative phase in the superposition \eqref{|w0s>}, one has that the marginal distributions are, in general, sensitive to this phase via the oscillatory term $\cos{(q_A-q_{0A})}$. Now, it is immediately seen from the factors $\Gamma_{q,cl}(n)$ that while the quantum distribution remains sensitive to the phase for arbitrarily long times (large $n$), the classical distribution loses its dependence on the phase as fast as $n^{-1}$. This stresses the limitations of our CHVM in following quantum mechanics to longer times. See Ref.~\cite{sanz03} for a related study indicating connections between entanglement oscillations and quantum interference in the Husimi representation.

Finally, it is worth noticing the role of the statistical mechanisms introduced in the model. Of course, they should be critical for the establishment of correlations. To deepen the discussion, let us consider the two-body correlation function
\be
\mathfrak{C}(O_A,O_B)=\langle O_A O_B\rangle-\langle O_A\rangle\langle O_B\rangle,
\ee 
where $O_{s}$ denote both quantum $(\hat{n}_{s}\cdot\vec{\sigma}^s)$ and classical $(\hat{n}_{s}\cdot\vec{\mathcal{J}}^s)$ spin observables defined by directions $\hat{n}_s$.
Calculations performed for $w_{0s}=1$ ($q_{0s}=p_{0s}=0$), which make the initial state reduce to an eigenstate of $\mathrm{J}_{A\mathrm{x}}\otimes \mathrm{J}_{B\mathrm{x}}$, give
\beq \label{frakC}
\!\!\!\!\!\!\mathfrak{C}_{\text{q,cl}}(\hat{n}_A,\hat{n}_B)\!=\!\left(\begin{matrix}n_{A\mathrm{x}} & n_{A\mathrm{y}} & n_{A\mathrm{z}}\end{matrix}\right)\left(\begin{matrix} u_{q,\text{cl}} & 0 & 0\\ 0 & 0 & v_{q,\text{cl}} \\ 0 & v_{q,\text{cl}} & 0\end{matrix}\right)\left(\begin{matrix}n_{B\mathrm{x}} \\ n_{B\mathrm{y}} \\ n_{B\mathrm{z}}\end{matrix}\right),\quad
\eeq
where 
\be\begin{array}{lll}
u_{q}=\tfrac{1}{4}\sin^2\tfrac{\tau}{2},& & u_{\text{cl}}=\tfrac{\delta^2}{\tau^4}\left[\cos^2\tau+\left(1-\frac{\tau^4}{9}\right)\frac{\sin^2\tau}{\tau^2}-\frac{\sin(2\tau)}{\tau}\right],\\ \\
v_{q}=\tfrac{1}{4}\sin\tfrac{\tau}{2},& & v_{\text{cl}}=\tfrac{\delta}{3\tau}\left(\frac{\sin\tau}{\tau}-\cos\tau\right).
\end{array}\nonumber \ee 
Clearly, there are infinitely many observables $(\hat{n}_A,\hat{n}_B)$ which will get correlated ($\mathfrak{C}>0$) as the time evolution starts. Also, except for $\tau=2\pi n$ $(n\in\mathbb{N})$ both quantum and classical quantities will agree about the presence of correlations, although not in their absolute values. Now, notice the presence of $\delta$ in all classical results appearing in Eqs.~\eqref{Cclresult}, \eqref{Hwp}, and \eqref{frakC}. In particular, observe that the classical formulas will dramatically fail to imitate the underlying quantum phenomena when $\delta=0$ but will perform better if $\delta=1$. The $\delta=0$ case corresponds to  a stationary regime in which the probability distribution reduces to a steady uniform plateau, $\wp(\vec{x},t)=\wp_0(\vec{x})=\tfrac{1}{4\pi}$. In fact, this distribution is an {\em absolute} steady state, as it neither evolves in time nor generates correlations regardless of the interaction model in question. In this framework, therefore, we may take the following interpretation: If, on the one hand, $\vec{x}$ plays the role of a hidden variable that implements {\em determinism} in our CHVM, on the other hand, $\delta$ is the hidden variable intended to introduce statistics (at an epistemic level) and emulate {\em nonlocal correlations}. As we have seen, however, $\delta$ fails in accomplishing its second task. 

\section{Concluding remarks}
\label{conclusion}

Aiming at getting a better understanding of the relation between entanglement and Bell nonlocality, we explicitly constructed a classical-hidden-variable model (CHVM) for the dynamics of a two-qubit system by adopting the usual classical limit of spin systems $(\hbar\to 0; j\to 1/\hbar)$. We employed the Liouville formalism to implement statistics in the model and proposed, in clear analogy with entanglement, a symmetrical measure of inseparability for the joint probability distribution. As a result, we showed that the measure of inseparability can furnish a good mimic for the entanglement generation between initially separable spin coherent states. Given the entanglement-nonlocality equivalence (Gisin's theorem), we then faced the question of whether the CHVM, provided with manifestly local hidden variables (the trajectories $\vec{x}$), would be at odds with Bell's theorem. To prove that this is not the case, we showed that the CHVM is not able to violate the CHSH inequality.

In other words, in the short-time regime, our model allows one to keep track of the entanglement dynamics but not directly of the nonlocality dynamics. This reveals that in the domain of pure states, violating a Bell inequality is much more demanding from a hidden-variable model than reproducing entanglement. As we have seen here for a two-qubit system and as has been reported elsewhere for other systems~\cite{angelo04,angelo05,matzkin11,casati12}, the short-time entanglement dynamics of initially separated coherent states seems to admit, in general, a reasonable mimic in terms of classical trajectories supplemented with statistical mechanisms; nonlocality does not. Thus, as the capacity of a hidden-variable model to describe quantum resources is taken as a figure of merit, we see that entanglement and nonlocality are not really equivalent, not even in the context of pure states. In principle, such a criterion can be applied to other quantum resources (e.g., Einstein-Podolsky-Rosen steering), and a hierarchy can be established. As a natural continuation of the present work, one may consider nonlocal interaction models [in the sense of producing nonlocal generalized forces, as in Eq.~\eqref{gen_force}]. It would be interesting to verify whether our CHVM would have a better chance to violate the CHSH inequality when provided with nonlocal trajectories.

\section*{Acknowledgments} We gratefully acknowledge inspiring discussions with K. Furuya ({\em in memoriam}), J. G. Peixoto de Faria, and A. D. Ribeiro. R.M.A. acknowledges financial support from the National Institute for Science and Technology of Quantum Information (INCT-IQ/CNPq, Brazil). 



\begin{thebibliography}{70}

\expandafter\ifx\csname natexlab\endcsname\relax\def\natexlab#1{#1}\fi
\expandafter\ifx\csname bibnamefont\endcsname\relax
  \def\bibnamefont#1{#1}\fi
\expandafter\ifx\csname bibfnamefont\endcsname\relax
  \def\bibfnamefont#1{#1}\fi
\expandafter\ifx\csname citenamefont\endcsname\relax
  \def\citenamefont#1{#1}\fi
\expandafter\ifx\csname url\endcsname\relax
  \def\url#1{\texttt{#1}}\fi
\expandafter\ifx\csname urlprefix\endcsname\relax\def\urlprefix{URL }\fi
\providecommand{\bibinfo}[2]{#2}
\providecommand{\eprint}[2][]{\url{#2}}

\bibitem[{\citenamefont{EPR}(1935)}]{EPR35}
\bibinfo{author}{\bibfnamefont{A.} \bibnamefont{Einstein}},
\bibinfo{author}{\bibfnamefont{B.} \bibnamefont{Podolsky}}, \bibnamefont{and}
\bibinfo{author}{\bibfnamefont{N.} \bibnamefont{Rosen}}, 
\bibinfo{author}{\bibfnamefont{Can quantum-mechanical description of physical reality be considered complete?}},
 \bibinfo{journal}{Phys. Rev.} \textbf{\bibinfo{volume}{47}},
 \bibinfo{pages}{777} (\bibinfo{year}{1935}). 
 
\bibitem[{\citenamefont{Bell}(1964)}]{bell64}
\bibinfo{author}{\bibfnamefont{J. S.} \bibnamefont{Bell}},
\bibinfo{author}{\bibfnamefont{On the Einstein-Podolsky-Rosen paradox}},
 \bibinfo{journal}{Physics} \textbf{\bibinfo{volume}{1}},
 \bibinfo{pages}{195} (\bibinfo{year}{1964}). 

\bibitem[{\citenamefont{Hensen et al}(2015)}]{hensen15}
\bibinfo{author}{\bibfnamefont{B.} \bibnamefont{Hensen}} \bibnamefont{{\it et al.}},
\bibinfo{author}{\bibfnamefont{Loophole-free Bell inequality violation using electron spins separated by 1.3 kilometres}},
 \bibinfo{journal}{Nature (London)} \textbf{\bibinfo{volume}{526}},
 \bibinfo{pages}{682} (\bibinfo{year}{2015}). 
 
\bibitem[{\citenamefont{Giustina et al}(2015)}]{giustina15}
\bibinfo{author}{\bibfnamefont{M.} \bibnamefont{Giustina}} \bibnamefont{{\it et al.}},
\bibinfo{author}{\bibfnamefont{Significant loophole-free test of Bell's theorem with entangled photons}},
 \bibinfo{journal}{Phys. Rev. Lett.} \textbf{\bibinfo{volume}{115}},
 \bibinfo{pages}{250401} (\bibinfo{year}{2015}). 

\bibitem[{\citenamefont{Shalm et al}(2015)}]{shalm15}
\bibinfo{author}{\bibfnamefont{L. K.} \bibnamefont{Shalm}} \bibnamefont{{\it et al.}},
\bibinfo{author}{\bibfnamefont{Strong loophole-free test of local realism}},
 \bibinfo{journal}{Phys. Rev. Lett.} \textbf{\bibinfo{volume}{115}},
 \bibinfo{pages}{250402} (\bibinfo{year}{2015}). 

\bibitem[{\citenamefont{Hensen et al}(2016)}]{hensen16}
\bibinfo{author}{\bibfnamefont{B.} \bibnamefont{Hensen}} \bibnamefont{{\it et al.}},
\bibinfo{author}{\bibfnamefont{Loophole-free Bell test using electron spins in diamond: Second experiment and additional analysis}},
 \bibinfo{journal}{Sci. Rep.} \textbf{\bibinfo{volume}{6}},
 \bibinfo{pages}{30289} (\bibinfo{year}{2016}). 
 
\bibitem[{\citenamefont{Horodecki et al}(2009)}]{horodecki09}
\bibinfo{author}{\bibfnamefont{R.} \bibnamefont{Horodecki}},
\bibinfo{author}{\bibfnamefont{P.} \bibnamefont{Horodecki}},
\bibinfo{author}{\bibfnamefont{M.} \bibnamefont{Horodecki}}, \bibnamefont{and}
\bibinfo{author}{\bibfnamefont{K.} \bibnamefont{Horodecki}}, 
\bibinfo{author}{\bibfnamefont{Quantum entanglement}},
 \bibinfo{journal}{Rev. Mod. Phys.} \textbf{\bibinfo{volume}{81}},
 \bibinfo{pages}{865} (\bibinfo{year}{2009}).
 
\bibitem[{\citenamefont{Wiseman et al}(2007)}]{wiseman07} 
\bibinfo{author}{\bibfnamefont{H. M.} \bibnamefont{Wiseman}}, 
\bibinfo{author}{\bibfnamefont{S. J.} \bibnamefont{Jones}}, \bibnamefont{and}
\bibinfo{author}{\bibfnamefont{A. C.} \bibnamefont{Doherty}},
\bibinfo{author}{\bibfnamefont{Steering, Entanglement, Nonlocality, and the Einstein-Podolsky-Rosen Paradox}},
  \bibinfo{journal}{Phys. Rev. Lett.} \textbf{\bibinfo{volume}{98}},
  \bibinfo{pages}{140402} (\bibinfo{year}{2007}).
  
\bibitem[{\citenamefont{Brunner et al}(2014)}]{brunner14}
\bibinfo{author}{\bibfnamefont{N.} \bibnamefont{Brunner}},
\bibinfo{author}{\bibfnamefont{D.} \bibnamefont{Cavalcanti}},
\bibinfo{author}{\bibfnamefont{S.} \bibnamefont{Pironio}},
\bibinfo{author}{\bibfnamefont{V.} \bibnamefont{Scarani}}, \bibnamefont{and}
\bibinfo{author}{\bibfnamefont{S.} \bibnamefont{Wehner}}, 
\bibinfo{author}{\bibfnamefont{Bell nonlocality,}}
 \bibinfo{journal}{Rev. Mod. Phys.} \textbf{\bibinfo{volume}{86}},
 \bibinfo{pages}{419} (\bibinfo{year}{2014}). 
  
\bibitem[{\citenamefont{Quintino et al}(2015)}]{quintino15}
\bibinfo{author}{\bibfnamefont{M. T.} \bibnamefont{Quintino}},
\bibinfo{author}{\bibfnamefont{T.} \bibnamefont{V\'ertesi}}, 
\bibinfo{author}{\bibfnamefont{D.} \bibnamefont{Cavalcanti}}, 
\bibinfo{author}{\bibfnamefont{R.} \bibnamefont{Augusiak}}, 
\bibinfo{author}{\bibfnamefont{M.} \bibnamefont{Demianowicz}},
\bibinfo{author}{\bibfnamefont{A.} \bibnamefont{Ac\'in}}, \bibnamefont{and}
\bibinfo{author}{\bibfnamefont{N.} \bibnamefont{Brunner}}, 
\bibinfo{author}{\bibfnamefont{Inequivalence of entanglement, steering, and Bell nonlocality for general measurements}},
  \bibinfo{journal}{Phys. Rev. A} \textbf{\bibinfo{volume}{92}},
  \bibinfo{pages}{032107} (\bibinfo{year}{2015}). 

\bibitem[{\citenamefont{Costa et al}(2016)}]{costa16}
\bibinfo{author}{\bibfnamefont{A. C. S.} \bibnamefont{Costa}},
\bibinfo{author}{\bibfnamefont{M. W.} \bibnamefont{Beims}}, \bibnamefont{and}
\bibinfo{author}{\bibfnamefont{R. M.} \bibnamefont{Angelo}}, 
\bibinfo{author}{\bibfnamefont{Generalized discord, entanglement, Einstein-Podolsky-Rosen steering, and Bell nonlocality in two-qubit systems under (non-)Markovian channels: Hierarchy of quantum resources and chronology of deaths and births}},
  \bibinfo{journal}{Phys. A (Amsterdam, Neth.)} \textbf{\bibinfo{volume}{461}},
  \bibinfo{pages}{469} (\bibinfo{year}{2016}). 

\bibitem[{\citenamefont{Bilobran}(2015)}]{bilobran15}
\bibinfo{author}{\bibfnamefont{A. L. O.} \bibnamefont{Bilobran}} \bibnamefont{and}
\bibinfo{author}{\bibfnamefont{R. M.} \bibnamefont{Angelo}},
\bibinfo{title}{\bibfnamefont{A measure of physical reality}},
  \bibinfo{journal}{Europhys. Lett.} \textbf{\bibinfo{volume}{112}}, 
  \bibinfo{pages}{40005} (\bibinfo{year}{2015}).  
 
\bibitem[{\citenamefont{Gisin}(1991)}]{gisin91}
\bibinfo{author}{\bibfnamefont{N.} \bibnamefont{Gisin}},
\bibinfo{title}{\bibfnamefont{Bell's inequality holds for all non-product states}},
  \bibinfo{journal}{Phys. Lett. A} \textbf{\bibinfo{volume}{154}}, 
  \bibinfo{pages}{201} (\bibinfo{year}{1991}).  
 
\bibitem[{\citenamefont{Yu et al}(2012)}]{yu12}
\bibinfo{author}{\bibfnamefont{S.} \bibnamefont{Yu}},
\bibinfo{author}{\bibfnamefont{Q.} \bibnamefont{Chen}},
\bibinfo{author}{\bibfnamefont{C.} \bibnamefont{Zhang}},
\bibinfo{author}{\bibfnamefont{C. H.} \bibnamefont{Lai}}, \bibnamefont{and}
\bibinfo{author}{\bibfnamefont{C. H.} \bibnamefont{Oh}},
\bibinfo{author}{\bibfnamefont{All entangled pure states violate a single Bell's inequality}},
 \bibinfo{journal}{Phys. Rev. Lett.} \textbf{\bibinfo{volume}{109}},
 \bibinfo{pages}{120402} (\bibinfo{year}{2012}). 
 
\bibitem[{\citenamefont{Bowles et al}(2016)}]{bowles16}
\bibinfo{author}{\bibfnamefont{J.} \bibnamefont{Bowles}},
\bibinfo{author}{\bibfnamefont{J.} \bibnamefont{Francfort}},
\bibinfo{author}{\bibfnamefont{M.} \bibnamefont{Fillettaz}},
\bibinfo{author}{\bibfnamefont{F.} \bibnamefont{Hirsch}},
 \bibnamefont{and}
\bibinfo{author}{\bibfnamefont{N.} \bibnamefont{Brunner}},
\bibinfo{author}{\bibfnamefont{Genuinely multipartite entangled quantum states with fully local hidden variable models and hidden multipartite nonlocality}},
 \bibinfo{journal}{Phys. Rev. Lett.} \textbf{\bibinfo{volume}{116}},
 \bibinfo{pages}{130401} (\bibinfo{year}{2016}). 

\bibitem[{\citenamefont{Cavalcanti et al}(2016)}]{cavalcanti16}
\bibinfo{author}{\bibfnamefont{D.} \bibnamefont{Cavalcanti}},
\bibinfo{author}{\bibfnamefont{L.} \bibnamefont{Guerini}},
\bibinfo{author}{\bibfnamefont{R.} \bibnamefont{Rabelo}}, \bibnamefont{and}
\bibinfo{author}{\bibfnamefont{P.} \bibnamefont{Skrzypczyk}},
\bibinfo{author}{\bibfnamefont{General method for constructing local-hidden-variable models for entangled quantum states}},
 \bibinfo{journal}{Phys. Rev. Lett.} \textbf{\bibinfo{volume}{117}},
 \bibinfo{pages}{190401} (\bibinfo{year}{2016}). 

\bibitem[{\citenamefont{Hirsch et al}(2016)}]{hirsch16}
\bibinfo{author}{\bibfnamefont{F.} \bibnamefont{Hirsch}},
\bibinfo{author}{\bibfnamefont{M. T.} \bibnamefont{Quintino}},
\bibinfo{author}{\bibfnamefont{T.} \bibnamefont{V\'ertesi}},
\bibinfo{author}{\bibfnamefont{M. F.} \bibnamefont{Pusey}},
 \bibnamefont{and}
\bibinfo{author}{\bibfnamefont{N.} \bibnamefont{Brunner}},
\bibinfo{author}{\bibfnamefont{Algorithmic construction of local hidden variable models for entangled quantum states}},
 \bibinfo{journal}{Phys. Rev. Lett.} \textbf{\bibinfo{volume}{117}},
 \bibinfo{pages}{190402} (\bibinfo{year}{2016}).

\bibitem[{\citenamefont{Angelo et al}(2004)}]{angelo04}
\bibinfo{author}{\bibfnamefont{R. M.} \bibnamefont{Angelo}},
\bibinfo{author}{\bibfnamefont{S. A.} \bibnamefont{Vitiello}},
\bibinfo{author}{\bibfnamefont{M. A. M.} \bibnamefont{de Aguiar}}, \bibnamefont{and}
\bibinfo{author}{\bibfnamefont{K.} \bibnamefont{Furuya}}, 
\bibinfo{author}{\bibfnamefont{Quantum linear mutual information and classical correlations in globally pure bipartite systems}},
 \bibinfo{journal}{Phys. A (Amsterdam, Neth.)} \textbf{\bibinfo{volume}{338}},
 \bibinfo{pages}{458} (\bibinfo{year}{2004}).
 
\bibitem[{\citenamefont{Angelo et al}(2005)}]{angelo05}
\bibinfo{author}{\bibfnamefont{R. M.} \bibnamefont{Angelo}} \bibnamefont{and}
\bibinfo{author}{\bibfnamefont{K.} \bibnamefont{Furuya}}, 
\bibinfo{author}{\bibfnamefont{Semiclassical limit of the entanglement in closed pure systems}},
 \bibinfo{journal}{Phys. Rev. A} \textbf{\bibinfo{volume}{71}},
 \bibinfo{pages}{042321} (\bibinfo{year}{2005}).

\bibitem[{\citenamefont{Matzkin}(2011)}]{matzkin11}
\bibinfo{author}{\bibfnamefont{A.} \bibnamefont{Matzkin}},
\bibinfo{author}{\bibfnamefont{Entanglement in the classical limit: Quantum correlations from classical probabilities}},
 \bibinfo{journal}{Phys. Rev. A} \textbf{\bibinfo{volume}{84}},
 \bibinfo{pages}{022111} (\bibinfo{year}{2011}).

\bibitem[{\citenamefont{Casati et al}(2012)}]{casati12}
\bibinfo{author}{\bibfnamefont{G.} \bibnamefont{Casati}},
\bibinfo{author}{\bibfnamefont{I.} \bibnamefont{Guarneri}}, \bibnamefont{and}
\bibinfo{author}{\bibfnamefont{J.} \bibnamefont{Reslen}}, 
\bibinfo{author}{\bibfnamefont{Classical dynamics of quantum entanglement}},
 \bibinfo{journal}{Phys. Rev. E} \textbf{\bibinfo{volume}{85}},
 \bibinfo{pages}{036208} (\bibinfo{year}{2012}).
 
\bibitem[{\citenamefont{Bohm}(1952)}]{bohm52}
\bibinfo{author}{\bibfnamefont{D.} \bibnamefont{Bohm}}, 
\bibinfo{author}{\bibfnamefont{A suggested interpretation of the quantum theory in terms of hidden
variables}},
 \bibinfo{journal}{Phys. Rev.} \textbf{\bibinfo{volume}{85}},
 \bibinfo{pages}{166} (\bibinfo{year}{1952}). 
 
\bibitem[{\citenamefont{Cavalcanti et al}(2012)}]{cavalcanti12}
\bibinfo{author}{\bibfnamefont{Eric G.} \bibnamefont{Cavalcanti}} \bibnamefont{and}
\bibinfo{author}{\bibfnamefont{Howard M.} \bibnamefont{Wiseman}}, 
\bibinfo{author}{\bibfnamefont{Bell nonlocality, signal locality and unpredictability
(or What Bohr could have told Einstein at Solvay had he known about Bell experiments)}},
 \bibinfo{journal}{Found. Phys.} \textbf{\bibinfo{volume}{42}},
 \bibinfo{pages}{1329} (\bibinfo{year}{2012}). 

\bibitem[{\citenamefont{Radcliffe}(1971)}]{radcliffe71}
\bibinfo{author}{\bibfnamefont{J. M.} \bibnamefont{Radcliffe}},
\bibinfo{author}{\bibfnamefont{Some properties of coherent spin states}},
 \bibinfo{journal}{J. Phys. A} \textbf{\bibinfo{volume}{4}},
 \bibinfo{pages}{3} (\bibinfo{year}{1971}).

\bibitem[{\citenamefont{Gazeau}(2009)}]{gazeau09}
\bibinfo{author}{\bibfnamefont{J.-P.} \bibnamefont{Gazeau}},
  \emph{\bibinfo{title}{Coherent States in Quantum Physics}}
  (\bibinfo{publisher}{Wiley-VCH, Weinheim, Germany}, \bibinfo{year}{2009}).

\bibitem[{\citenamefont{Dowling et al}(1994)}]{dowling94}
\bibinfo{author}{\bibfnamefont{J. P.} \bibnamefont{Dowling}},
\bibinfo{author}{\bibfnamefont{G. S.} \bibnamefont{Agarwal}}, \bibnamefont{and}
\bibinfo{author}{\bibfnamefont{W. P.} \bibnamefont{Schleich}},
\bibinfo{author}{\bibfnamefont{Wigner distribution of a general angular-momentum state: Applications to a collection of two-level atoms}},
 \bibinfo{journal}{Phys. Rev. A} \textbf{\bibinfo{volume}{49}},
 \bibinfo{pages}{4101} (\bibinfo{year}{1994}).
 
\bibitem[{\citenamefont{Sanz et al}(2003)}]{sanz03}
\bibinfo{author}{\bibfnamefont{L.} \bibnamefont{Sanz}},
\bibinfo{author}{\bibfnamefont{R. M.} \bibnamefont{Angelo}}, \bibnamefont{and}
\bibinfo{author}{\bibfnamefont{K.} \bibnamefont{Furuya}},
\bibinfo{author}{\bibfnamefont{Entanglement dynamics in a two-mode nonlinear bosonic Hamiltonian}},
 \bibinfo{journal}{J. Phys. A} \textbf{\bibinfo{volume}{36}},
 \bibinfo{pages}{9737} (\bibinfo{year}{2003}).
 
\end{thebibliography}
\end{document}